\documentclass[twocolumn,showpacs,preprintnumbers,amsmath,amssymb]{revtex4}

\usepackage{graphicx}
\usepackage{dcolumn}
\usepackage{bm}

\begin{document}

\preprint{APS/123-QED}

\title{Dynamics of crater formations in immersed granular materials}

\author{Germ\'an Varas, Val\'erie Vidal and Jean-Christophe G\'eminard}
\affiliation{Laboratoire de Physique, Universit\'e de Lyon, Ecole Normale Sup\'erieure - CNRS,
46 All\'ee d'Italie, 69364 Lyon Cedex, France}

\date{\today}        

\begin{abstract}
We report the formation of a crater at the free surface of an immersed granular bed,
locally crossed by an ascending gas flow.
In two dimensions, the crater consists of two piles which develop around the location of the gas emission. 
We observe that the typical size of the crater increases logarithmically with time,
independently of the gas emission dynamics.
We describe the related granular flows and give an account of the influence of the
experimental parameters, especially of the grain size and of the gas flow.
\end{abstract}

\pacs{83.80.Fg : Granular materials, rheology,
47.57.Gc : Granular flow, complex fluid,
47.85.Dh : Hydrodynamics, applied fluid mechanics}

\maketitle

\section{\label{sec:level1}Introduction}

Craters are part of the widespread phenomena observed in nature,
going from impact meteorite craters to volcanic structures.
Studies of crater morphologies have a wide range of applications, going from puzzling
crater formation in drying paint \cite{Evans00} to molecular dynamics 
\cite{Insepov03,Yamaguchi03,Aoki05}. 
Among the main applications to natural phenomena, aside from meteorite impact 
crater, are the formation and growth of volcanic edifices, by successive ejecta
emplacement and/or erosion. The time evolution and dynamics play a crucial role here, 
as the competition between volcanic-jet mass-flux (degassing and ejecta)
and crater-size evolution may control directly the eruptive regime \cite{Woods95}. 
Attempts have been made to model the talus development of volcanic caldera through
erosion \cite{Obanawa08}, or to constrain the morphology and dynamics of pyroclastic 
constructs via granular-heap drainage laboratory experiments \cite{Riedel03}.

Crater morphology in dry granular material has been extensively studied,
both experimentally and theoretically \cite{Grasselli01,Uehara03,Lohse04,Zheng04,Wada06,deVet07}. 
Most of these studies investigate the final, steady, crater shape resulting from the collision
of solid bodies with the material surface and scaling laws are derived \cite{Walsh03}.
Note however that some authors also reported experimental study of the crater formation dynamics,
including growth and collapse after impact, in the "single impacting body" configuration
\cite{Boudet06,Yamamoto06}.
In a recent work, Wu {\it et al.} have extended these studies to a particles stream
impacting a dry granular bed \cite{Wu07}. 

In immersed granular material, one reports craters generated by an underwater vortex ring \cite{Suzuki07},
involving fluidized ejecta dynamics, or underwater impact craters generated by landslide \cite{Fritz03}.
Craters in immersed granular materials can result either from two-phase or three-phase flows. 
In particular, water or gas flowing through an immersed granular bed can induce
localized instabilities and fluidization \cite{Rigord05,Zoueshtiagh07},
which eventually leads to the formation of craters at the free surface.

In a previous experimental study, Gostiaux \emph{et al.} \cite{Gostiaux02} have investigated
the dynamics of air flowing through an immersed granular layer.
They reported that, depending on the flow rate, the system exhibits two qualitatively
different regimes:
At small flow rate, the bubbling regime during which bubbles escape the granular
layer quite independently one from another;
At large flow rate, the open-channel regime which corresponds to the formation 
of a channel crossing the whole thickness of the granular bed through which air escapes
almost continuously. At intermediate flow rate, a spontaneous alternation between these
two regimes is observed. 
Interestingly, they noticed the appearance of a crater around the locus of air release
but did not provide any extensive study of its formation dynamics.
Here, we extend these seminal observations to a quantitative study of the 
resulting deformation of the free surface of the granular bed.
To do so, we reproduce the previous experimental conditions in two dimensions: 
In a vertical Hele-Shaw cell, the crater is then formed by two sand piles which grow and move
away from each other as time passes by. By monitoring the evolution ot the free surface through time, 
we investigate the effect of the different gas flow regimes on the crater dynamics. 
The results point out a grain-transport mechanism that differs significantly from the one involved 
in dry-sand dune motion \cite{Bagnold,Kroy,Andreotti,Hersen}.

\section{Experimental setup and protocol}
\label{sec:expsetup}

\begin{figure*}
\includegraphics[width=1.8\columnwidth]{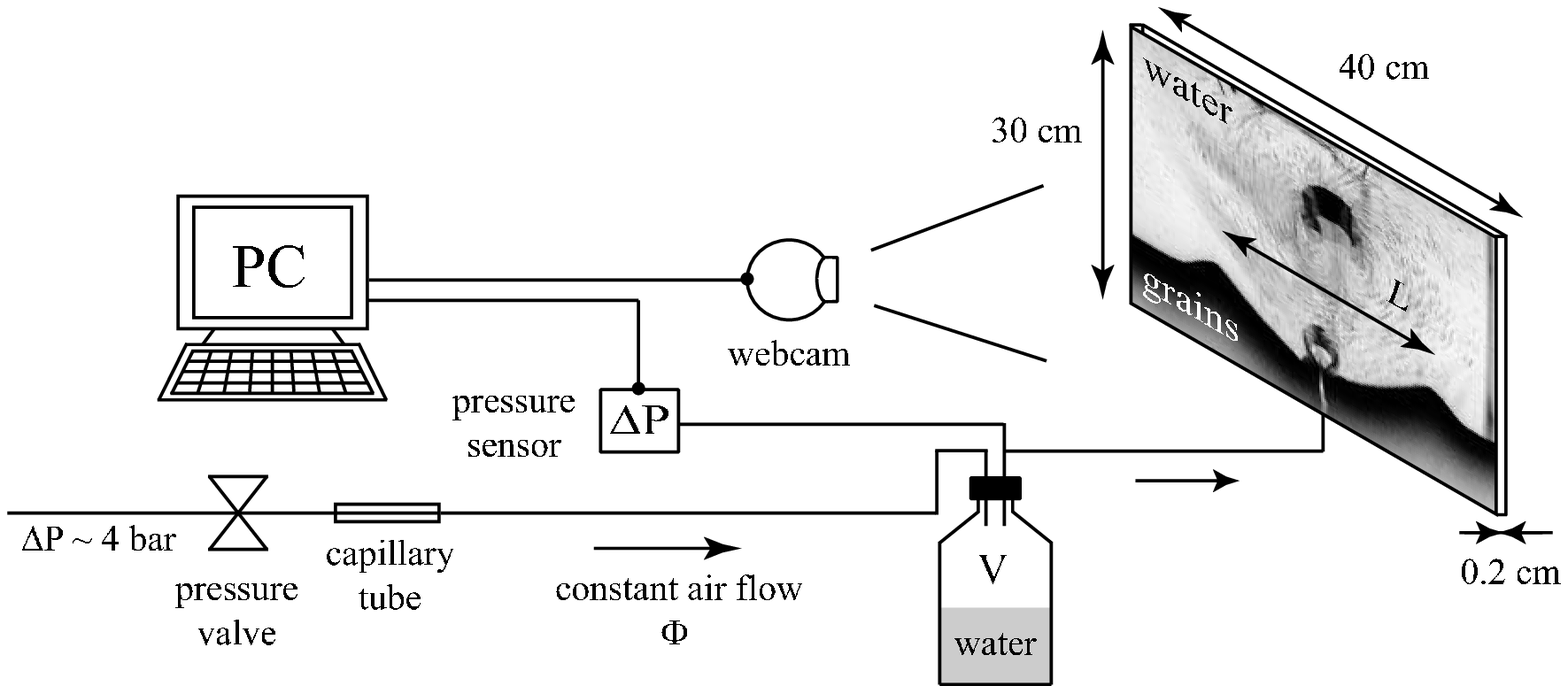}
\caption{\label{fig:expsetup} {\bf Experimental setup:} Air is injected at constant
flow rate $\Phi$ into a chamber (volume $V$) connected to a thin vertical cell containing
an immersed, initially flat, granular layer.
Through time, symmetrical piles form, grow and move in opposite direction on either side of the crater.
The experiment consists in observing the deformation of the granular-bed free-surface and in monitoring
the overpressure $\Delta P$ inside the chamber, in order to account for the influence of the degassing
process on the crater dynamics  [the picture here illustrates the bubbling regime (see text).]}
\end{figure*}

The principle of the experiment is to force air, injected locally at the base of an immersed granular
bed, to flow vertically across the material.
The experimental setup thus consists in a 2-mm-thin vertical-cell containing the grains and water 
(Fig.~\ref{fig:expsetup}). The two vertical walls (glass plates) are 40~cm wide and 30~cm high. 

The air injection is assured by a system similar to that used in \cite{Gostiaux02}:
Air is injected, at a constant flow rate, in a chamber connected to the bottom of the cell.
In practice, a small hole (diameter 1~mm) connects,
at the center, the bottom of the vertical cell to the chamber, partially filled
with water in order to tune the inner gaseous volume $V$.
The air flow is controlled thanks to a reducing valve which 
maintains a constant overpressure at the entrance of a capillary tube:
Provided that the pressure drop in the capillary tube is much larger than the fluctuations of the pressure
inside the chamber, the air flow toward the chamber is not significantly altered by the dynamics of the granular layer
and remains constant.
The flow rate $\Phi$ (ranging from 1.5~mL/s to 4.3~mL/s) is tuned by changing either the capillary tube (rough) or
the pressure difference imposed by the reducing valve (fine) and is subsequently measured 
(to within 0.1~mL/s) by means of a home-made flow meter.
A sensor  (MKS Instruments, 223 BD-00010 AB), 
connected to an acquisition board (National Instruments, PCI-6251), monitors the pressure 
difference, $\Delta P$ (to within 10~Pa), between the chamber and the outside atmosphere.

In order to account for the dynamics of the free surface, we image 
the system from the side.
In a first configuration, a transparency flat viewer (Just NormLicht,  Classic Line)
located behind the cell is used to achieve a homogeneous lighting of the whole bed
which is imaged with the help of a Webcam (Logitech, QuickCam Express) connected to a PC.
A small software (Astra Image Webcam Video Grabber) makes it possible to take 
one image of the system every 10 seconds and, thus, to record the dynamics
during several hours (typically 24 hours). 
In a second configuration, in order to observe the motion of the grains, 
we light up the sample from top with a linear light source (Polytec, DCR3)
and take pictures with a high-resolution digital-camera (Nikon, D200).

Initially, $\Phi$ is set to zero. Grains and water are introduced in the cell.
The grains consist of spherical glass-beads 
(USF Matrasur, sodosilicate glass) that are sieved in order to control their size 
(typical diameters, 100--125 or 400--500~$\mu$m). Initially, after sedimentation, the grains 
sit at the bottom and we make use of a thin rod to level the immersed granular layer. 
We denote $h_g$ the height of grains above the bottom of the container 
and $h_w$, that of the water free surface above the granular bed.  

The initial condition consists thus in a horizontal and flat layer of grains.
After the opening of the valve, one observes that air is creating several paths
between the hole at the bottom and the free surface of the granular bed.
During this transient regime, some air channels merge, or cannot reach the upper
layer of grains (In this case, air bubbles remain trapped within the granular layer.)
After several minutes, air crosses the granular bed along the vertical and bubbles
are emitted in the water from an almost fixed position. 
Then, through time, a crater forms, the two piles on both sides growing and moving apart
one from the other (see Fig.~\ref{fig:expsetup}).

\section{Results}
\label{sec:results}

We shall report the dynamics of the crater formation in regard to the gas flow regimes.
First, we shall describe qualitatively the gas flow regimes (Sec.~\ref{sec:degassing})
and the associated mechanisms of the gas emission at the free surface (Sec.~\ref{sec:mechanisms}).
Then, we shall describe qualitatively the crater formation (Sec.~\ref{sec:geometry})
and make use of these first observations to estimate the profile of the grain deposition.
We shall also discuss the effects of the finite depth of the granular bed and of the
finite water height (Sec.~\ref{sec:finite}).
We finally quantitatively study the influence of the gas flux (Sec.~\ref{sec:flux})
and of the grain size (Sec.~\ref{sec:grains}).

\subsection{Qualitative observations.}
\label{sec:qualitative observations}

\subsubsection{Gas flow regimes}
\label{sec:degassing}

\begin{figure}[t]
\includegraphics[width=\columnwidth]{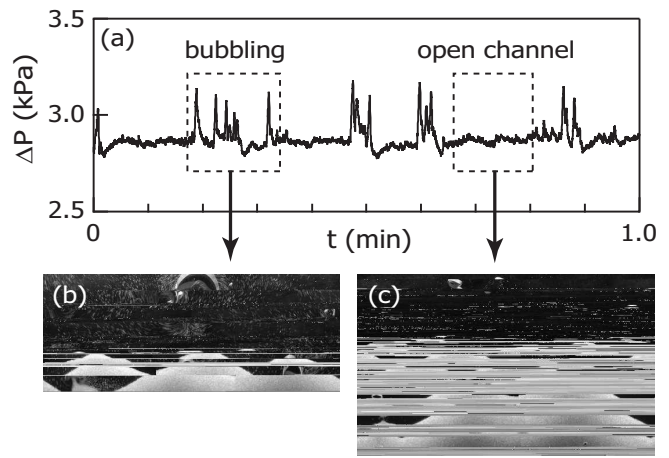}
\caption{\label{fig:pressure} {\bf Overview on the gas flow process:}
(a) At intermediate flow-rate $\Phi$, the pressure signal exhibits a spontaneous
alternation between activity and rest periods which correspond respectively to the
bubbling and open-channel regimes.
The emission of the bubbles at the free surface accordingly exhibits two
qualitatively different behaviors : (b) In the bubbling regime,
a large bubble forms underneath the free surface of the granular bed and then bursts.
The explosion pushes out of the bed a large number of grains, which are subsequently
advected upwards in the bubble wake.
(c) In the open-channel regime, the walls of the channel are stable and a continuous
gas flow escapes the granular bed. In this regime the grains are only torn out from the granular
bed by the water flow and subsequently advected in the wake of the small bubbles that form in water
($d = $ 400~$\mu$m, $h_g=8$~cm, $h_{w}=16$~cm, and $\Phi=2.5$~mL/s).}
\end{figure}

As already reported \cite{Gostiaux02}, depending on the flow rate $\Phi$,
two main regimes of the air flowing through the granular layer are observed.
(The existence of the two regimes and the transition between them are tightly related to
the non-Newtonian rheology of the granular material as proven by an experimental study
conducted in another non-Newtonian material \cite{Divoux08}.)
On the one hand, the bubbling regime, which is typically observed at small $\Phi$,
is characterized by a regular emission of successive bubbles, independent from one another.
In this regime, the pressure signal exhibits successive rises and drops, the latter being
associated with bubble emissions from the injector at the base of the granular layer.
On the other hand, in the open-channel regime, which is typically observed at large $\Phi$,
the system sustains a continuous air-flow through a channel crossing the whole granular layer.
The overpressure $\Delta P$ associated with this continuous air emission is almost constant.
In an intermediate range of $\Phi$, one observes a spontaneous alternation between
the two regimes: The channel forms after the emission of several bubbles and
subsequently spontaneously pinches off after a finite time, leading the
system back to the bubbling regime. 
As a consequence, activity and rest periods are observed in the pressure signal [Fig.~\ref{fig:pressure}(a)].
The phenomenon is explained by the ability of the material to sustain a stable channel thanks to its
peculiar rheology.
At last, we point out that the deformation of the free surface of the granular bed does
not seem to alter the gas flow process: It has been previously shown that the gas emission 
is mainly governed by local events occuring close to the free surface \cite{Gostiaux02}.

\subsubsection{Bubble emission mechanisms}
\label{sec:mechanisms}

A close look at the free surface of the granular bed points out two different air-release mechanisms
associated with the bubbling and open-channel regimes, respectively.
In the bubbling regime, a gas bubble, while growing underneath the free surface, pushes up a thin layer of grains
which forms the bubble head.
Once it has crossed the interface, the bubble, while it rises up in the water,
advects the grains in its wake [Fig.~\ref{fig:pressure}(b)]. 
In the open-channel regime, the air is released continuously through the channel whose walls remain
at rest. The grains are advected upwards, from the free surface, by the water flow behind the small ascending
gas bubbles [Fig.~\ref{fig:pressure}(c)].
From these observations, one could wonder about the relative efficiency of the two regimes in forming the crater.
In particular, one could expect the bubbling regime to lead to a faster growth
because the explosive bursting of the bubble apparently lifts a larger quantity of grains.
In Sec. \ref{sec:parameters}, in order to answer the question, we shall report a quantitative study
of the crater growth in large range of $\Phi$.

\subsubsection{Geometrical description of the crater}
\label{sec:geometry}

\begin{figure}[t]
\includegraphics[height=\columnwidth,angle=-90]{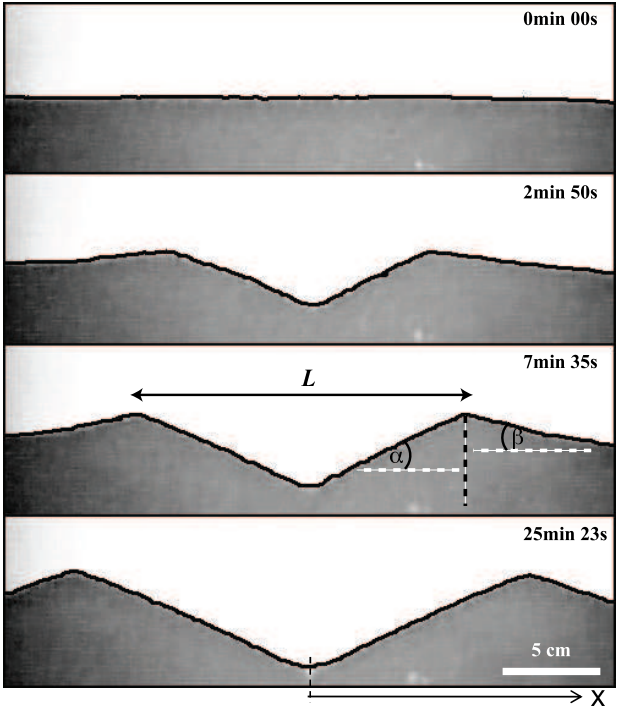}
\caption{\label{fig:talus} {\bf Temporal evolution of the crater:} 
The black lines are the result of the automatic detection of the free-surface profile. 
The angles $\alpha$ and $\beta$ denote, respectively, the maximum angle the inner and outer
flanks make with the horizontal. We define $L$, the distance between the two peaks,
and $x$, the distance from the center
($d = 100~\mu$m, $h_g=8$~cm, $h_w=16$~cm, and $\Phi=2.2$~mL/s).}
\end{figure}

The ejection of the grains from the free surface and the subsequent
deposition of the granular ejecta lead to the rapid formation of a crater,
which is formed by two granular piles symmetrically positioned on each flank.
Due to the permanent grain transport, the crater grows:
The piles height increases while they move symmetrically away
from the center (Fig.~\ref{fig:talus}).

\begin{figure}[h]
\includegraphics[width=\columnwidth]{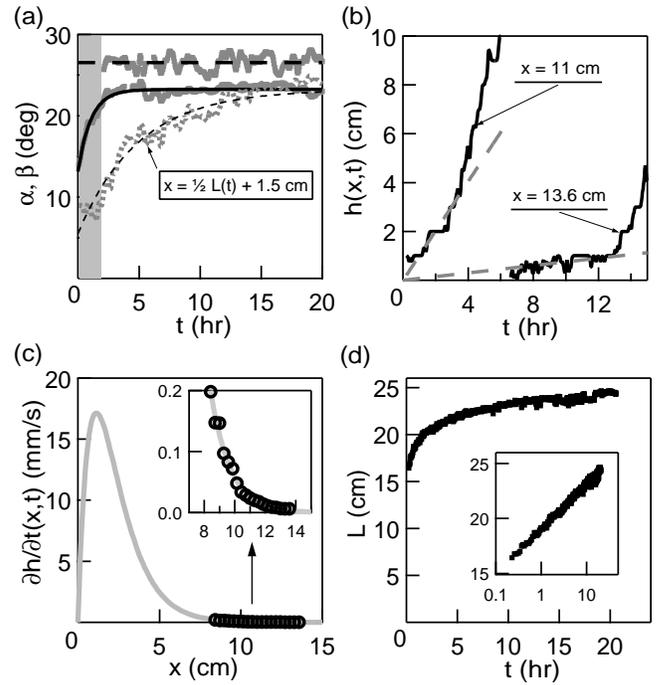}
\caption{\label{fig:data} {\bf Temporal evolution of the crater geometry:} 
(a) Inner and outer angles $\alpha$ and $\beta$ vs time $t$ ($\Phi=1.67$~mL/s).
The grey zone marks the early stages of the crater growth during which
the crater size compares with the size of the bubbles. The angle $\alpha$,
which is then not measurable, remains constant during the whole experimental time.
In contrast, the angle $\beta$ increases continuously to reach a constant value
after a finite time. In order to show that the slope is not constant along the outer flanks,
we report the temporal evolution of the slope at $x = \frac{1}{2}L(t) + 1.5$~cm: the local slope
remains smaller than $\beta$ until reached by the avalanche flow. Lines are only guides for the eye.
(b) Local height $h(x,t)$ vs time $t$ ($\Phi=2.93$~mL/s). Far away from the center, the height
$h$ of the granular layer increases due to the deposition of the grains.
We observe a linear increase of $h(x,t)$ with time until the region is reached
by the grains avalanching along the outer flank. 
(c) Local growth velocity $\partial h/ \partial t(x,t)$ vs distance $x$ ($\Phi=2.93$~mL/s).
The data are successfully interpolated by $a \frac{x}{L_c} \exp{(-\frac{x}{L_c})}$
for large $x$, we get at a rough estimate $L_c = (1.13 \pm 0.04)$~cm.
(d) Distance $L$ between the summits vs time $t$ ($\Phi=4.26$~mL/s). Whereas the crater forms
quite fast, we observe a drastic decrease of the growth velocity at large times.
Inset : $L$ vs $\ln(t)$. At large times, $L$ increases almost logarithmically with time $t$
($d = 400~\mu$m, $h_g=8$~cm, and $h_w=16$~cm).}
\end{figure}

Before reporting quantitative measurements of the crater dynamics,
let us mention some qualitative features of the crater formation.
First, let us denote $\alpha$ and $\beta$ the maximum angles that
the piles make with the horizontal, respectively, inside and outside the
crater. From direct observation [Fig.~\ref{fig:talus}], one can notice
immediately that $\alpha$ and $\beta$ can differ significantly, especially
at the early stages of the crater formation [Fig.~\ref{fig:data}(a)].
We note that the inner flanks of the crater are almost
straight and that $\alpha$ (about 28$^\circ$)
remains constant during the whole crater growth.
In contrast, the outer slopes are not straight, especially at the early stages
of the crater formation, and $\beta$ evolves in time.

\begin{figure}[!tbp]
\includegraphics[width=\columnwidth]{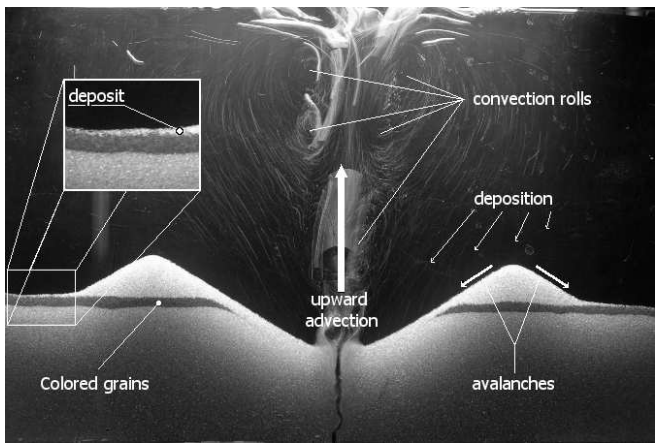}
\caption{\label{fig:flows} {\bf The granular flows:} 
The second configuration is used: 
we light up the sample from top with the linear light source
and take pictures with the high-resolution digital camera.
Initially, a thin layer of colored grains is deposited at the free
surface of the initially flat and horizontal bed. Then, a series
of ascending rolls pushes the grains away from the vertical central axis.
Subsequently, the grains gently deposit back onto the free surface 
of the bed. Along the piles flanks, provided that the local angle
exceeds the angle of avalanche, we observe continuous avalanches.
Inside the crater, the flowing granular-material partly replaces
the grains advected upwards at the center. Along the outer flanks,
the deposited granular material either flows or sits at the free 
surface. One can clearly observe, in the inset, that new material,
deposited far away from the center forms a thin layer of material
which remains at rest. The deposition flow rate, $Q(x)$,
is estimated from the temporal evolution of the
growth velocity $\partial h/\partial t$ of the deposited layer in that region  [Fig.~\ref{fig:data}(c)]
($d = 400~\mu$m, $h_g=8$~cm, $h_w=16$~cm, and $\Phi = 3.5~$mL/s).}
\end{figure}

This behavior can be qualitatively
accounted for by considering the grain flows (Fig.~\ref{fig:flows}).
Once lifted up by liquid flow, the grains are pushed away from the center
by the liquid flow which mainly consists of two large convective rolls and
then, subjected to gravity, deposit back onto the free surface at finite 
distance from the center. If the local slope is smaller than the angle of
avalanche the grains do not move anymore once deposited. To the contrary, if
the local slope is larger than the angle of avalanche, the grains flow
downwards along the slope.
At the center the grains are very locally torn off from the surface by the upward liquid flow.
The local slope almost immediately exceeds the angle of avalanche and grains flow downwards
on both inner flanks to replace the granular material missing at the center. 
Thus, if a grain deposits onto the inner flank,
it flows downwards toward the center, which again explains why $\alpha$
remains almost constant and only fluctuates between 
the angle of repose and the angle of avalanche [Fig.~\ref{fig:data}(a)].
In constrast, at the early stages of the crater formation, the outer flanks only result
from the deposition of the granular material. 
The local slope of the outer flank being everywhere smaller than
the angle of avalanche, a grain, once deposited, does not move anymore. 
However, at a finite time $t$, $\beta$ reaches the value of the angle of avalanche,
which results in surface flows along the outer flank. As a consequence,
$\beta$ evolves in time [Fig.~\ref{fig:data}(a)] :
Starting from a small value at the early stages of the crater formation,
$\beta$ increases to reach a constant value after a finite time.
We note here that $\beta$ then remains slightly smaller than $\alpha$,
which is probably explained by the fact that $\alpha$ significantly
exceeds the angle of avalanche due to the continuous grain flow along the 
inner flanks.

At this point, it is particularly interesting to make use of these observations
to estimate the spatial distribution of the deposited grains. To do so, let us now
consider the local height $h(x,t)$ of the free surface at the distance $x$ from the
center at time $t$ [Fig.~\ref{fig:data}(b)]. Close to the center, $h(x,t)$ exhibits
a complex behavior which results from both the deposition and the surface flows.
Far from the center, $h(x,t)$, which results from the deposition alone, evolves
linearly in time. From this latter observation, we deduce that the advection 
is not significantly affected by the crater growth and that the distribution
of the deposited grains, far away from the center, is almost constant in time.
At an intermediate distance from the center, we observe a change in the temporal
evolution of $h(x,t)$ : At small times, the dynamics is only due to the deposition
whereas later, when $\beta$ reaches the avalanche angle, the local dynamics results
from both the deposition and the surface flows. The volume, $Q(x)$, of granular 
material deposited at the distance $x$ from the center per unit distance and per unit time
is proportional to the initial slope $\partial h(x,t)/\partial t$ far away from the center
[Fig.~\ref{fig:data}(c)].
Seeking for a simple mathematical description of the deposition flow, we guess that 
the grains, pushed away from the center by the liquid flow, have a negligible probability
to deposit back at the center and we propose to interpolate $\partial h(x,t)/\partial t$  
by $a \frac{x}{L_c} \exp{(-\frac{x}{L_c})}$, which makes it possible to extract a characteristic
length $L_c$ at which the grains are deposited away from the center.
We shall show in Sec. \ref{sec:parameters} that the shape of the distribution $Q(x)$
accounts for the evolution of the distance $L(t)$ between the two summits as a function
of the time $t$ [Fig.~\ref{fig:data}(d)].  

\subsubsection{Finite size effects}
\label{sec:finite}

We aim at reporting the growth of the crater in an infinite surround,
not limited by the finite depth of the granular bed, $h_g$, or by the
finite water height, $h_w$.

The growth of the crater stops when the thickness of the granular bed at the center vanishes.
Taking into account the angle of avalanche and the conservation of the granular-material volume,
we can estimate the maximum accessible value of $L$,
$L_{max}^g = (2+\sqrt{2}) h_g / \tan{\alpha}$.
We shall report experimental results obtained for $h_g = 8$~cm
so that $L_{max} \sim 50$~cm is about the lateral width
of the cell. The growth of the crater shall thus not be limited
by the finite depth of the granular bed.

\begin{figure}[h]
\includegraphics[width=0.9\columnwidth]{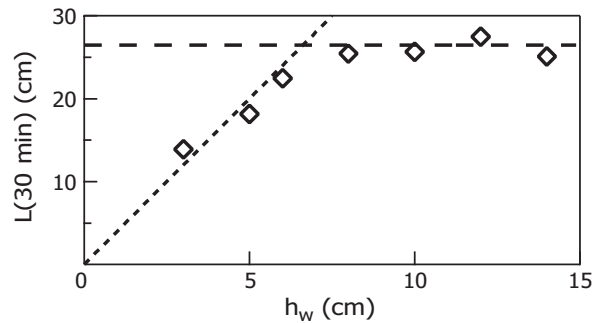}
\caption{\label{fig:eau} {\bf Distance $L(30\,{\rm min})$ vs water height $h_w$:}
For $h_w < L/4$, the growth of the crater is limited
by the water height and $L(30\,{\rm min})$ increases linearly with $h_w$.
In contrast, for $h_w > L/4$, the growth of the crater is 
not altered by the water height and $L(30\,{\rm min})$ does not depend on $h_w$
($d~=~100~\mu$m, $h_g~=~8$~cm, and $\Phi~=~3.06$~mL/s).}
\end{figure}
In the same manner, the height of the piles can obviously not exceed $h_w$.
We estimate, the maximum accessible value of $L$,
$L_{max}^w = 2(1+\sqrt{2}) h_w / \tan{\alpha}$.
However, we checked experimentally that $L_{max}^w$
largely underestimates the finite water-height effect:
Reporting $L$ at a given, large, time $t=30$~min 
as a function of $h_w$, we obtain experimentally
that the growth of the crater is limited by the 
water height for $h_w < L/4$ (Fig.~\ref{fig:eau}).
Interestingly, we observe, in addition, that 
$L(30\,{\rm min})$ does not depend on $h_w$ if $h_w > L/4$ :
The crater growth is not altered by the water height if the latter is large enough.
We shall report experimental results obtained for $h_w = 16$~cm
so that $L_{max}^w \sim 64$~cm is larger than the lateral width
of the cell. The growth of the crater shall thus not be limited
by the finite water height.

\subsection{Influence of the gas flow $\Phi$ and grain size $d$}
\label{sec:parameters}

\subsubsection{Dependence on the air flux $\Phi$}
\label{sec:flux}

\begin{figure}
\includegraphics[width=\columnwidth]{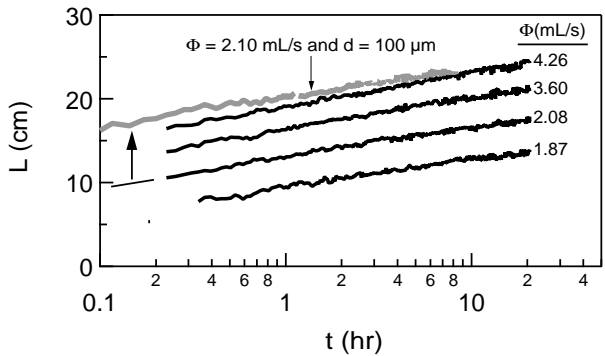}
\caption{\label{fig:distance} {\bf Distance $L$ vs. time $t$:}
The distance $L$ between the summits increases logarithmically with time, 
independently from the gas emission regimes. 
Indeed, at small $\Phi$, the gas flow regime consists mainly in the 
emission of independent bubbles, whereas it consists mainly in a continuous 
emission of gas at large $\Phi$. 
From the interpolation of the experimental data with $L/L_0 = \ln (\Phi t / V_0)$,
we estimate $L_0 = (3.5\pm0.7)$~cm and $V_0 = (0.4\pm0.2)$~mL for $d = 400~\mu$m (black curves)
whereas we estimate $L_0 = (3.5\pm1)$~cm and $V_0 = (0.02\pm0.01)$~mL for $d = 100~\mu$m (grey curve)
($h_g=8$~cm, $h_w=16$~cm).}
\end{figure}
We report the distance $L$ between the piles
as a function of time $t$, for different air flow rates $\Phi$ (Fig.~\ref{fig:distance}).
We observe that $L$ increases logarithmically with $t$ and we write
$L/L_0 = \ln (\Phi t / V_0)$, $V_0$ being a typical volume.

In the bubbling regime, when the gas emission at the free surface consists of
the periodic emission of independent bubbles at the free surface, a simple argument can
account for the proposed dependance on the total emitted gas volume, $\Phi t$, alone:
The dimensions of the piles, length $L$ or height,  are expected to depend on the
total number $N_g$ of grains displaced by the water flow.
Thus, $N_g$ being proportional to the number of emitted bubbles,
which itself, at a given time $t$, is proportional to $\Phi t$, we expect $L$ to be 
a function of $\Phi t$ alone, as observed experimentally.
In addition, we observe that the relation holds true even at large $\Phi$ when the
system exhibits, almost always, the open-channel regime.
Thus, contrary to the intuition, the small bubbles emitted by the open-channel
are as efficient as the exploding bubble of the bubbling regime in lifting the grains:
This observation is again in agreement with the fact that the grains are mainly lifted
by the water flow and not expelled by the explosion of the bubbles \cite{Gostiaux02}.   

\subsubsection{Dependence on the grain size $d$}
\label{sec:grains}

From the interpolation of the exprimental data, one can estimate 
the typical length $L_0$ and the typical volume $V_0$.
One observes that the length $L_0$ does not depend significantly on the grain size
(the slope in the semilog plot remains almost unchanged) and we
estimate $L_0 = (3.5\pm1)$~cm for $d = 400~\mu$m and $d = 100~\mu$m.
In contrast, we estimate $V_0 \sim 0.4~$mL for $d = 400~\mu$m
and a significantly smaller value $V_0 \sim 0.02~$mL for $d = 100~\mu$m.
The precise dependence of $L_0$ and $V_0$ on $d$ is difficult to 
access experimentally as such measurements require precise knowledge of the
origin of time (a delay alters significantly the slope and the offset in the
semilog plot) which we are missing. Nevertheless, $V_0$ is observed to
decrease drastically with the grain size whereas $L_0$ remains almost constant.

\section{Discussion}
\label{sec:model}

The crater growth is due to the advection of the grains and to the subsequent
deposition away from the location of the gas emission.
From the observation of the deposit far away from the center (Sec.~\ref{sec:geometry}), 
we propose that the deposited flow can be written, for large distance $x$,
$Q(x) = a \frac{x}{L_c} \exp{(-\frac{x}{L_c})}$.
This relation accounts for the logarithmic growth of the crater.
Indeed, assuming the angle $\alpha$ on both sides of the piles,
a simple geometrical analysis gives the volume, $v = \frac{1}{8} \frac{\tan \alpha}{(1+\sqrt{2})^2} L^2$
of the grains on the outer side of one pile ($x \ge L/2$). 
Noting that only the deposition of grains on the outer flank of the pile
contributes to its growth, we can write $\frac{dv}{dt} = \int_{L/2}^\infty Q(x)\,dx$,
which leads to the equation governing the pile growth,
\begin{equation}
\frac{1}{4} \frac{\tan \alpha}{(1+\sqrt{2})^2} L \frac{dL}{dt} = a \int_{L/2}^\infty  \frac{x}{L_c} \exp{\left(-\frac{x}{L_c}\right)dx}.
\label{croissance}
\end{equation}
Provided that $L \gg L_c$ (which is almost always satisfied experimentally),
we obtain that, asymptotically, $L$ increases logarithmically with time $t$
according to $L(t) \simeq L_0 \ln{({\Phi t}/{V_0})}$
with $L_0 = 2 L_c$ and $\Phi/V_0=a{(1+\sqrt{2})^2}/{\tan \alpha}$.
The experimental rough estimates, obtained for $d = 400~\mu$m, $L_c = (1.13\pm0.04)$~cm
(Sec.~\ref{sec:geometry}) and $L_0 = (3.5\pm0.7)$~cm (Sec.~\ref{sec:flux}) are in fair agreement
with this expectation. The same conclusion holds true for the smaller grains, $d = 100~\mu$m,
as we measured $L_c = (1.38\pm0.04)$~cm for the same value of $L_0$.

Finally, if we interpret the volume $V_0$ as the gas volume necessary to lift or move a given
quantity of grains, one can easily understand that $V_0$ decreases with $d$.
Indeed, the advection process, whose associated force scales like $d$, competes
with the buoyancy force which scales like $d^3$. As a result, large grains are more
difficult to lift than small ones and the gaseous volume necessary to move
them away is larger.

The advection process remains difficult to model in details. Indeed, the liquid flow
in the wake of the bubbles is generally turbulent and it is barely possible to account
for the advection of the solid particles in such a complex stream field. There also remains
a very important open question: We observed that the growth of the crater is not altered
by the water depth (if large enough). However, we observe that the vertical size of the 
large-scale convective rolls compares to $h_w$, whereas their lateral size compares to $L/2$.
One thus would expect the typical distance $L_c$ to depend on $h_w$ and, even, on time.
Our study does not display such an effect. The typical vertical distance to 
be taken into account is probably the distance over which a grain follows the bubble
in its wake, in other words, the maximum accessed altitude.  

In conclusion, we reported experimental data on the dynamics of a crater growth at the free surface of
an immersed granular crossed by an upward gas flow. We observed that, due to the peculiar
transport of the grains in the wake of the rising bubbles, the typical size of the
crater increases logarithmically with time. The dynamics is demonstrated not to 
be altered by the gas flow regime and only to depend on the overall gas flow rate
and the typical size of the grains. However, even if we clearly established that the global dynamics of the 
crater is compatible with the spatial structure of the grain deposition around the gas-emission {\it locus},
we are still missing a complete modeling of the grain advection by the turbulent flow,
which deserves to be further investigated.

\begin{center}
{\bf ACKNOWLEDGMENT}
\end{center}

G.V. acknowledges a grant by CONICYT (Comisi\'on Nacional de Investigaci\'on 
Cient\'ifica y Tecnol\'ogica, Gobierno de Chile).

\end{document}